\documentstyle[twoside,fleqn,epsfig,espcrc2]{article}
\title{Search for neutrinoless tau decays  $\tau\rightarrow 3 \ell$ and $\tau\rightarrow \ell K^0_{S}$}

\author{Y.~Yusa$^{a}$, H.~Hayashii$^{b}$, T.~Nagamine$^{a}$, A.~Yamaguchi$^{a}$ for the Belle collaboration\\
$^{a}$ Department of Physics, Tohoku University,\\
Aramaki, Aoba-ku, Sendai 980-8578, Japan.\\
$^{b}$ Department of Physics, Nara Women's University,\\
Kita-Uoya, Nishi-machi, Nara 630-8506, Japan.\\
}
       
\begin{document}

\begin{abstract}
Neutrinoless  tau-lepton decays into either three leptons ($\tau^{-}\rightarrow \ell_{1}^{-} \ell_{2} \ell_{3}$) or one lepton and one $K_S^0$ meson($\tau^{-}\rightarrow \ell^{-} K^0_{S}$) where lepton $\ell$ means either an electron or muon, have been searched for using 48.6 fb$^{-1}$ of data collected with the Belle detector at the KEKB  $e^+e^-$ collider. No evidence for candidate decays are found in any channel. Therefore we set 90\% confidence level upper limits on the branching fraction for 8 different decay modes. These limits are more stringent than those set previously and reach to the $10^{-7}$ level.

\vspace{1pc}
\end{abstract}

\maketitle

{\renewcommand{\thefootnote}{\fnsymbol{footnote}}

\normalsize

\setcounter{footnote}{0}

\normalsize
\section{Introduction}

The recent observation of neutrino oscillations by Super-Kamiokande\cite{SK} and SNO\cite{SNO} suggests non-zero neutrino masses and flavor mixing in the lepton sector. This implies that charged lepton flavor violating processes such as $\tau^{-}\rightarrow \mu^{-}\gamma$, $\tau^{-}\rightarrow\mu^{-}\mu\mu$, $\tau^{-}\rightarrow \mu^{-} K_{S}^0$ also occur at some level. It is therefore important to search for lepton flavor violation in  rare decays of charged leptons.

 In the standard model(SM), lepton flavor violating(LFV) decays of the charged lepton are highly suppressed even if we consider the effect of neutrino mixing\cite{SMLFV}. However, many extensions of the Standard Model predict LFV decays of changed leptons \cite{LFVREV}. For example, in the minimum SUSY standard model with right-handed neutrinos\cite{SUSYNR1,SUSYNR2} or in a SUSY grand unified theory model\cite{SUSYGUT}, flavor mixing in the slepton mass matrix at the TeV scale can induce a LFV transition $\tau \rightarrow \mu(e)$.
The predicted rate of $\tau \to \mu \gamma$ decays in these models depends on the model parameters but ranges from $O(10^{-8}$) to $O(10^{-6})$ , which is interested area of the current and future $B$-factory experiments. On the other hand, other models, with multi-Higgs\cite{MHIGGS}, extra $Z^{'}$ gauge bosons  \cite{ZPRIM}, R-parity violating interaction\cite{RVIO} or generic models involving heavy neutrinos\cite{HNL,HNLILA},  predict  relatively large branching fractions for $\tau^- \rightarrow \ell_{1}^{-}\ell_{2}\ell_{3}$  decays. The maximum allowed value can reach the  $O(10^{-7})$ - $O(10^{-6})$ level. Searches for flavor violating decays in various decay modes are thus important and can provide strong constraints on possible new physics.\\

 Our recent results for the decay $\tau^{-}\rightarrow \mu^{-}\gamma$ are reported in another paper\cite{mugamma}. In this paper, we present the results of a search for  LFV decays into three leptons or one lepton and one $K_{S}^{0}$ meson;
 \begin{eqnarray}
 \quad\quad   \tau^{-} &\rightarrow&  \ell_{1}^{-} \ell_{2} \ell_{3}  \\
 \quad\quad   \tau^{-} &\rightarrow& \ell^{-} K^0_{S}         
 \end{eqnarray}
where $\ell_{i}$ stands for either an electron or muon\cite{foot}.

 In previous experiments, the most stringent upper limit  on the branching fraction for the decay  $\tau^{-}\rightarrow \ell_{1}^{-}\ell_{2}\ell_{3}$ has been set by the CLEO experiment. They obtained an upper limits of (1-2) $\times 10^{-6}$ at 90\% confidence level from a data sample of 4.79 fb$^{-1}$\cite{CLEO3L}. For the decays $\tau^{-}\rightarrow \ell^{-} K^0_{S}$,  previously published upper limits come from the Mark II Collaboration and are of $O(10^{-4})$\cite{MARKII}.  The CLEO group has recently reported the improved upper limits  $B(\tau^{-}\rightarrow e^{-}K^{0}_{S})< 9.1 \times 10^{-7}$, $B(\tau^{-}\rightarrow \mu^{-}K^{0}_{S})< 9.5 \times 10^{-7}$  at 90\% C.L. using a 13.9 fb$^{-1}$ data sample\cite{CLEOLKS}.

\section{Data and Event Selection}

 The data used in this analysis  were collected with the Belle detector\cite{Belle} at the KEKB\cite{KEKB}  asymmetric energy $e^{+}e^{-}$ collider operating at a  center-of-mass energy  $\sqrt{s}\sim$ 10.6 GeV. Among the recorded samples, we use 48.6 fb$^{-1}$ of data, corresponding to 44.2 million  $\tau$-pair productions, taken in the period between 1999 and 2001.

  The Belle detector is a general purpose detector having excellent capabilities for precise vertex determination as well as the particle identification. 
Tracking of charged particles is performed by a three layer double-sided silicon vertex detector (SVD) and a 50 layer cylindrical drift chamber (CDC) within the 1.5 T magnetic field. Charged hadron are identified by means of $dE/dx$ from the CDC, signal pulse-heights from the aerogel $\check{\rm C}$erenkov counters (ACC), and timing information from the time-of-flight scintillation counters (TOF). Muons are detected by a 14 layers of resistive plate counter interleaved with iron plates (KLM). Photons and electrons are detected using a CsI(Tl) electromagnetic calorimeter (ECL).\\

 In order to determine the selection criteria, a Monte Carlo (MC) studies are performed using the KORALB/TAUOLA program~\cite{KORALB} for the $\tau$-pair generation, the QQ generator\cite{QQ} for $B\bar{B}$ and $q\bar{q}$ continuum processes, the AAFHB\cite{AAFHB} program for  two-photon processes, the BHLUMI\cite{BHLUMI} program for radiative Bhabha events and the KKMC\cite{KKMC} program for radiative $\mu$-pair events. 

 Regarding the signal $\tau$-pair sample, one $\tau$ decays into $\ell_1 \ell_2 \ell_3$ or $\ell K^0$ assuming 3- or 2-body phase-space distribution, respectively, in the rest frame of $\tau$, and the other $\tau$ decays generically. The detector response is simulated by a GEANT\cite{GEANT} based program. The MC events thus prepared are passed through the same reconstruction and analysis program  as for the actual data. Based on the study of simulated signal and background samples, the candidate signal events are required to satisfy the criteria described below.

\subsection{Topological Event Selection}
\vspace*{0.2cm}

 To attain a large detection efficiency, we require 1-prong decay ($B(\tau \to$ ``{\it 1}-{\it prong}'') = 85.4\%) for one $\tau$ lepton in the reaction $e^+e^-\rightarrow \tau^+ \tau^-$ and the other $\tau$ decays into three charged final state.  Accordingly the signal candidate events must contain four charged tracks with zero net charge.  The tracks should be well-reconstructed and have transverse momentum $P_{t} > 0.1$ GeV/$c$ within a polar angle 17$^\circ < \theta_{lab} < 150^\circ$ in the laboratory frame.  The tracks include the ones originating from the beam interaction regions as well as the daughter tracks from the secondary $V^{0}$ vertex. The latter tracks are added in order to include the decay products of the $K_{S}^{0}$ meson.  To reject beam-gas events, the distance of closest approach of each non-$K_{S}^{0}$ track to the interaction point must be within $\pm 1$ cm transversely and $\pm 5$ cm along the beam. Photons are defined as neutral ECL clusters with an energy $E_\gamma >$ 0.1 GeV. Photons must be separated by at least 30 cm from projection of any charged tracks on the surface of the calorimeter.\\
   
 We divide each event into two hemispheres in the $e^+e^-$ CM system with the plane perpendicular to the thrust axis. The thrust axis is calculated from the momenta of charged tracks and photons. In each hemisphere, one(tag) side must contain one charged track. The other(signal) hemisphere must contain three identified leptons or one identified lepton and one $K_{S}^0$ meson. There is no explicit cut on the maximum number of photons on both hemispheres in order to maintain high detection efficiency while minimizing the dependence on MC simulation of showers.

 In order to reject the two-photon processes, total transverse momentum in the event is required to be larger than 0.5 and 0.2 GeV/$c$ for $\tau \to \ell_{1}^- \ell_2 \ell_3$ and $\tau \to \ell K^0_S$ decays, respectively. To reduce the radiative Bhabha and radiative $\mu$-pair background, the collinearity angle $\theta_{col}$ between signal and 1-prong side is required to satisfy $\theta_{col} <175^{\circ}$.

\subsection{Lepton Identification} 
\vspace*{0.2cm}
 Lepton identification is substantial in this analysis. Electrons are identified by means of a likelihood ratio combining  information on the ratio of the cluster energy in the ECL to the track momentum measured in the CDC and $dE/dx$ in the CDC. The electron identification efficiency is estimated to be 92\% for the momentum range between 1.0 GeV/$c$ and 3.0 GeV/$c$ in the laboratory frame. The rate for misidentifying hadrons as electrons is quite small (0.3\%). In order to correct for the energy loss from the radiation in the  material of the SVD region, the energy for the electron candidate is recalculated by adding the energy of radiated photon clusters, if an ECL cluster with energy less than 1.0 GeV is detected within a cone angle of 10$^\circ$ around the flight direction of the electron candidate track.

Muon are required to have a well reconstructed track in the muon system, comprised of 14 layers of iron plates interleaved with the KLM. The muon probability is calculated from two variables: the difference between the range calculated by the momentum of a particle and the range measured by the KLM as well as the $\chi^2$ of the KLM hits with respect to the extrapolated track. The identification efficiency is estimated to be $\sim$91\% for momenta in the laboratory frame greater than 1.2 GeV/$c$. The  rate  for misidentifying hadrons as muons is 2.0\%.  

\subsection{$K^0_S$ Selection}
\vspace*{0.2cm}
 $K_S^{0}$ candidate is selected from  $V^{0}$ candidates assuming their decay products are charged pions. The $V^{0}$ candidates are reconstructed from  all pairs of oppositely charged tracks with the following requirements; \\
 i)  the closest distance between two helices of the\\
\hspace*{0.3cm} candidate pair must be less than 1 cm in the\\
\hspace*{0.3cm}  coordinate along the beam axis,\\
ii) the flight length from the beam interaction\\
\hspace*{0.3cm} point to the crossing point in the transverse\\
\hspace*{0.3cm} plane must  exceed 2 mm, \\
iii) the angle between the direction of crossing\\
\hspace*{0.3cm} point and the direction of the $V^{0}$ momentum\\
\hspace*{0.3cm} must be less than 10$^\circ$.\\

Figure 1 shows the $\pi^+\pi^-$ invariant mass distribution for the data after applying topological conditions for events which include four charged tracks. A clean $K_S^{0}$ signal is seen with very low level background. The solid curve in the figure is a fit with a double Gaussian for the signal and a linear function for background. The fitted mean of the Gaussian fit is consistent with the nominal $K^0_S$  mass. The mass resolution for the $K_S^{0}$ signal is $ 5.1\pm 0.1$ MeV. The $K_S^{0}$ candidates must have an invariant mass $M(\pi^+\pi^-)$  within three standard deviations of the nominal $K_S^{0}$ mass, 0.485  $<M_{\pi^+\pi^-}<$ 0.510 GeV/$c^2$. 

\subsection{Signal Selection}
\vspace*{0.2cm}
To identify  the neutrinoless tau decays,  a signal region is defined by using two kinematical variables;
\begin{eqnarray}
\Delta E^* &\equiv& E^*_{signal} - E^{*}_{beam}  \quad   \nonumber \\
 \Delta M &\equiv& M_{signal} - M_{\tau},            \nonumber
\end{eqnarray} 
where $E^{*}_{beam}$  and $M_\tau$ are the beam energy in the $e^+e^-$ CM system  and the $\tau$ lepton mass, respectively. $E^*_{signal}$  and $M_{signal}$ are the sum of the  energy in the signal side and its invariant mass, respectively.  $E^*_{signal}$ is measured in the $e^+e^-$ CM system.
 Figure \ref{EvsMsmc} shows the $\Delta E^*$ vs. $\Delta M$ plots for the signal Monte Carlo sample. 
The signals are concentrated in the vicinity of  $\Delta E^*$ = 0  and $\Delta M$ = 0. The tail of $\Delta E^*$  on the lower side is caused by initial photon radiation in the $\tau$-pair production process. A small tail, seen on lower side of $\Delta M$ for the processes involving electrons, is caused by the leakage of electron showers in ECL. The signal region for the LFV decay modes
\begin{eqnarray}
-0.29  < \Delta E^* < 0.11 ~{\rm GeV} \nonumber \\
-25 < \Delta M  < 25 ~{\rm MeV}/c^2, \nonumber 
\end{eqnarray} 
is shown by the dashed lines in Fig. \ref{EvsMsmc}. These requirements correspond to three standard deviation limits.
 Figure \ref{EvsMdata} shows the $\Delta E^*$ vs. $\Delta M$ plots for the experimental data after applying all selection cuts except for $\Delta E^*$ and $\Delta M$. Each decay mode includes the decay products of the $\tau^-$ lepton as well as the charge conjugate decay mode from the $\tau^+$ lepton.

\subsection{Background and Efficiency}
\vspace*{0.2cm}
 The backgrounds remaining in the sample can be estimated from the side-band region shown in Fig.\ref{EvsMdata}. The background is negligible for the modes with $\tau$ decays into three leptons, while somewhat larger background is observed for the decays $\tau^- \to \ell^- K_S^{0}$. Study of the background MC indicates that this background comes mainly from the  decays $\tau^- \to \pi^-(K^-) K_S^{0}\nu_{\tau}$, where  a charged pion or kaon is misidentified as an electron or muon.

 In order to estimate the background in the signal box, a fit with a linear function is carried out for the $\Delta E^*$ distribution of the side-band region for the sample -25 $< \Delta M <$ 25 MeV/$c^2$. Since the statistics is low, the slope of the function in the fit is fixed to the one determined  by the fit of the same distribution but selected with looser or no  lepton identification requirements
 \footnote{
  Since the background are mainly from the fake lepton identification, its shape is expected to be similar in all cases.}.
The results for the estimated backgrounds are shown in Table 1. The error on the background comes from uncertainty in the fitting procedure.

 The signal detection efficiency is determined from the signal MC. The efficiencies for each of the LFV decay modes are listed in Table 1. The efficiency is about 10\% for $\tau^{-} \to \ell_1^{-} \ell_2 \ell_3$ decays, and is about 16\% for $\tau^{-} \to \ell^{-} K^0_S$ decays. The latter value includes the branching fraction for $K^0_S \to \pi^+ \pi^-$ ($68.60\pm 0.27\%$) as well as the branching fraction of the 1-prong decay of the tau lepton.\\

 The systematic errors for the efficiency determination come from tracking (2\%/track), $K_{S}^{0}$ selection (0.4\%), electron identification (1.5\%) and muon identification (1.3\%). The uncertainty in the luminosity measurement is estimated to be 1.5\%. In the calculation of the upper limit on the branching fraction, these systematics are taken into account by reducing the efficiencies by $1\sigma$ from its normal MC values.

\section{Results}

 No candidate decays are observed in the signal box for any of the eight decay modes, as can be seen in Fig. \ref{EvsMdata} and Table 1.

 Using the Feldman and Cousins method \cite{Poisson-limit,PDG} for zero observed events and the estimated number of background events in the signal box, upper limits are calculated for each decay mode. The final results are summarized in Table 1 for the 8 decay modes.

The upper limits on the branching fractions are 
\begin{eqnarray} 
B(\tau^{-} \to e^{-}e^{+}e^{-})  &<& 2.7\times 10^{-7} \nonumber \\
B(\tau^{-} \to \mu^{-}\mu^{+}\mu^{-}) &<& 3.8\times 10^{-7} \nonumber \\
B(\tau^{-} \to e^{-}\mu^{+}\mu^{-}) &<& 3.1\times 10^{-7} \nonumber \\
B(\tau^{-} \to \mu^{-}e^{+}e^{-}) &<& 2.4\times 10^{-7} \nonumber \\
B(\tau^{-} \to e^{+}\mu^{-}\mu^{-})& <& 3.2\times 10^{-7} \nonumber \\
B(\tau^{-} \to \mu^{+}e^{-}e^{-}) &<& 2.8\times 10^{-7} \nonumber \\
B(\tau^{-} \to e^{-} K_S^{0})  &<& 2.9\times 10^{-7} \nonumber \\
B(\tau^{-} \to \mu^{-} K_S^{0})  &<& 2.7\times 10^{-7} \nonumber 
\end{eqnarray}
at 90\% confidence level.

The limits obtained in this analysis for the decays $\tau^{-} \to \ell^{-}_{1} \ell_{2} \ell_{3}$ are approximately one order of magnitude more restrictive than the current limits. The upper limits for  $\tau^{-} \to \ell^{-} K^0_S$ decays are a factor of 4-5 more stringent than recently reported by the CLEO experiment\cite{CLEOLKS}.
 
Since the background level is still low, we expect to be able to search for the LFV decay mode at down to the $ O(10^{-8})$ level in the near future at high luminosity $B$-factory experiments.

\vspace*{20cm}

\newpage

\begin{table*}
\caption{Summary of detection efficiency, observed events, expected background and 90\% C.L. upper limits on the branching fractions.}
 \begin{tabular}{lccccr}
  \hline
  \hline
     &                  & Observed & Expected  ~~ & ~Upper limit & Upper limit\\
Mode & Efficiency(\%) & events   & background ~~& ~(This Exp.) & (PDG2002)  \\
  \hline
$\tau^- \to e^- e^+ e^-$       & $12.1 \pm 0.7$ & 0 & 0.4 $\pm$ 0.3 &  2.7 $\times$ 10$^{-7}$ & 2.9 $\times$ 10$^{-6}$\\
  \hline
$\tau^- \to \mu^- \mu^+ \mu^-$ & $9.0 \pm 0.6$ & 0 & 0.0 $\pm$ 0.1 &  3.8 $\times$ 10$^{-7}$ & 1.9 $\times$ 10$^{-6}$\\
  \hline
$\tau^- \to e^- \mu^+ \mu^-$   & $10.4 \pm 0.6$ & 0 & 0.5 $\pm$ 0.4 &  3.1 $\times$ 10$^{-7}$ & 1.8 $\times$ 10$^{-6}$\\
  \hline
$\tau^- \to \mu^- e^+ e^-$     & $11.9 \pm 0.7$ & 0 & 0.8 $\pm$ 0.4 &  2.4 $\times$ 10$^{-7}$ & 1.7 $\times$ 10$^{-6}$\\
  \hline
$\tau^- \to e^+ \mu^- \mu^-$   & $10.8 \pm 0.6$ & 0 & 0.0 $\pm$ 0.0 &  3.2 $\times$ 10$^{-7}$ & 1.5 $\times$ 10$^{-6}$\\
  \hline
$\tau^- \to \mu^+ e^- e^-$     & $12.2 \pm 0.7$ & 0 & 0.0 $\pm$ 0.0 &  2.8 $\times$ 10$^{-7}$ & 1.5 $\times$ 10$^{-6}$\\
  \hline
$\tau^- \to e^- K^0_S$           & $16.0 \pm 0.7$ & 0 & 1.7 $\pm$ 0.5 &  2.9 $\times$ 10$^{-7}$ & 1.3 $\times$ 10$^{-3}$\\
  \hline
$\tau^- \to \mu^- K^0_S$         & $17.4 \pm 0.7$ & 0 & 2.0 $\pm$ 0.5 &  2.7 $\times$ 10$^{-7}$ & 1.0 $\times$ 10$^{-3}$\\
  \hline
  \hline
 \end{tabular}
\end{table*}

\begin{figure*}[ht!]
\label{fig:Ksmass}
\begin{center}
\includegraphics[height=8cm,width=16cm,clip]{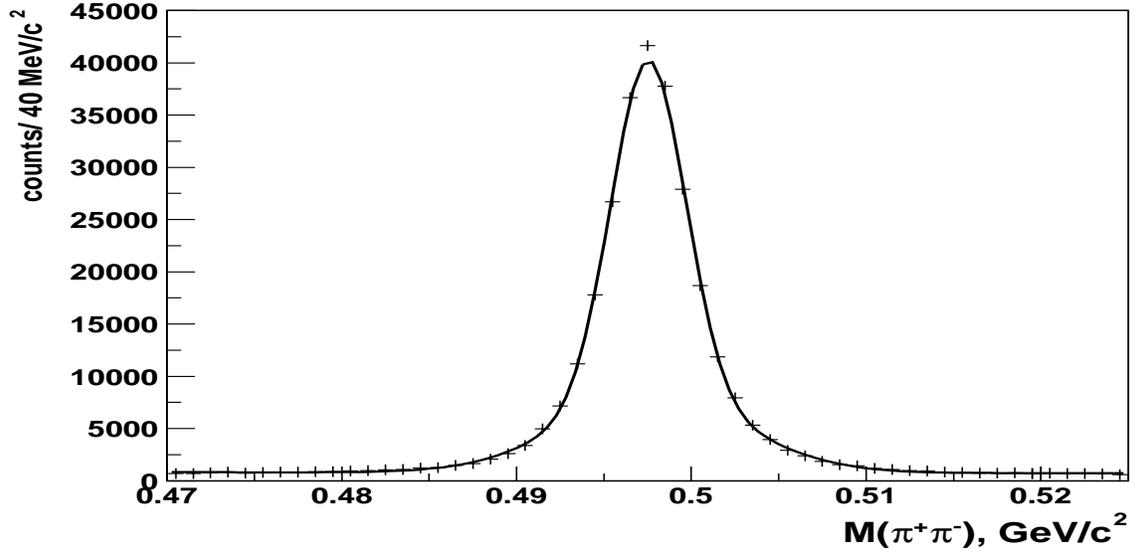}
\caption{
$\pi^+\pi^-$ invariant mass distribution for the experimental data sample. The solid line is the result
of a fit with a Gaussian for the K$^0_S$ signal and a linear function for
the background.
}
\end{center}
\end{figure*}

\newpage

\begin{figure*}[ht!]
\begin{center}
\vspace*{-1.5cm}
\includegraphics[width=170mm, height=80mm,clip]{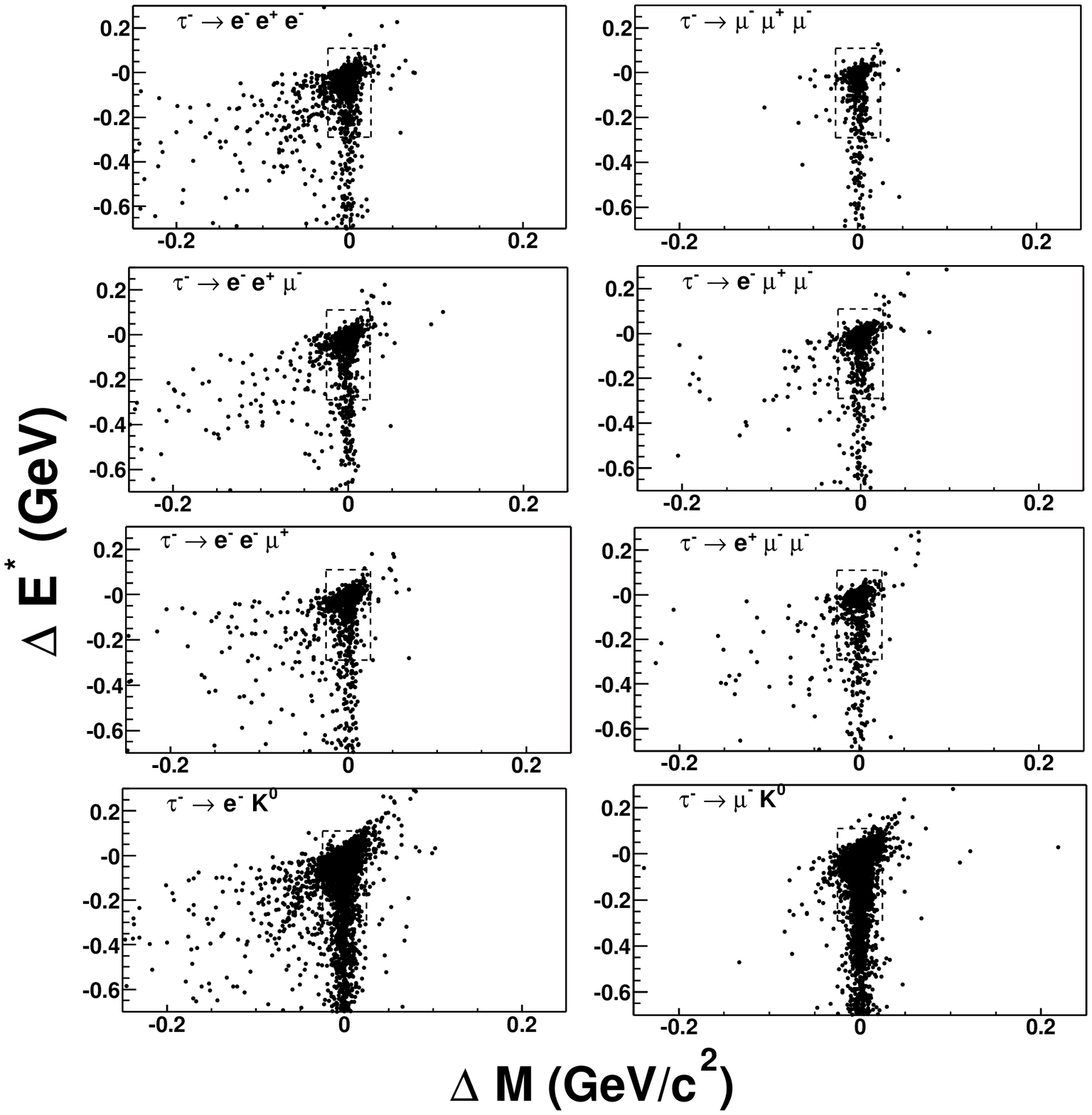}
\caption{
$\Delta E^*$ vs $\Delta M$ plots for the LFV decay modes in the signal Monte Carlo sample.
The dashed lines indicate the signal region.
}
\label{EvsMsmc}

\includegraphics[width=170mm, height=80mm,clip]{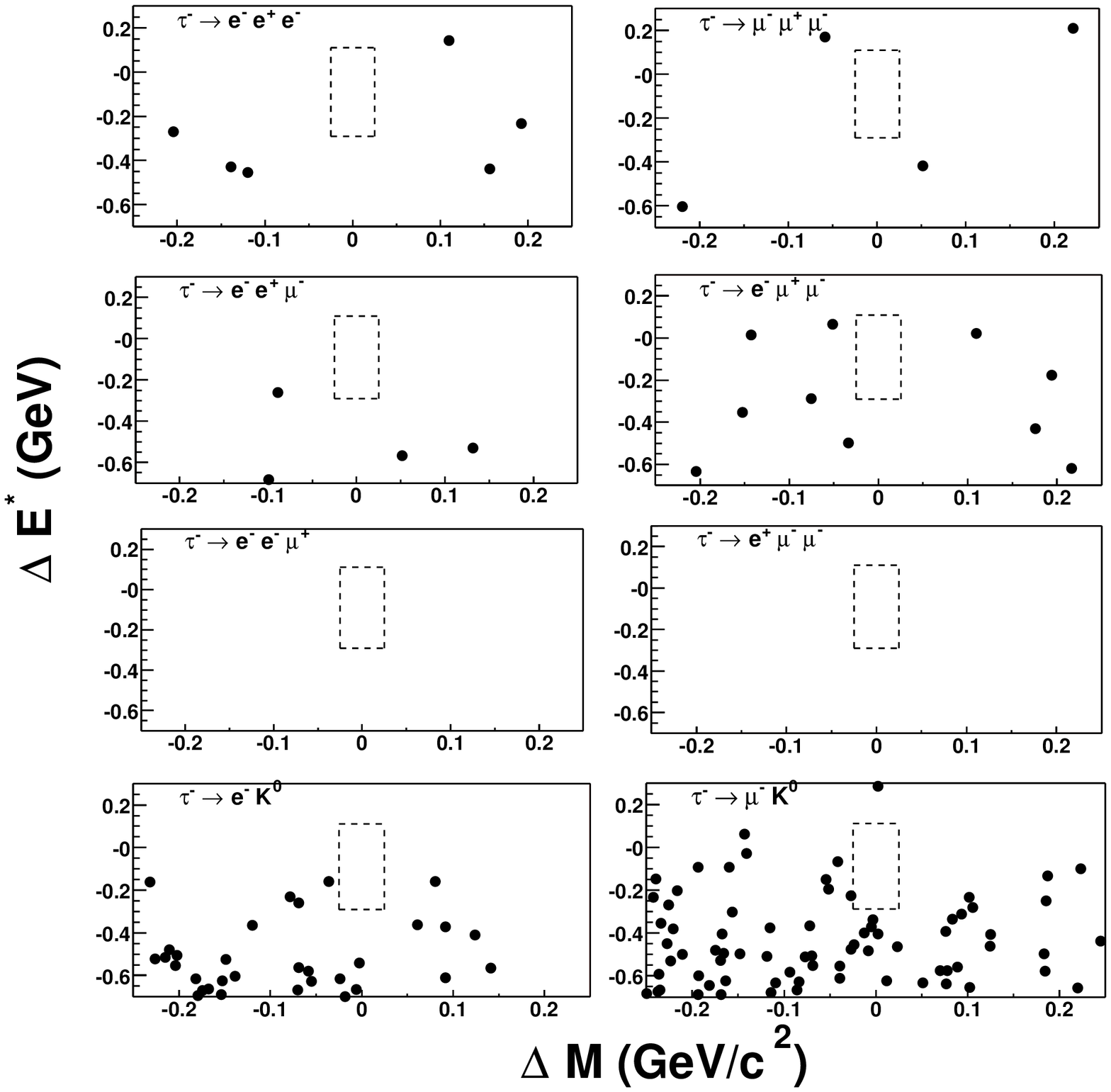}
\caption{
$\Delta E^*$ vs $\Delta M$ plots for the
experimental data sample.
The charge conjugate decay mode is also includes.
The dashed lines indicate the signal region.
}
\label{EvsMdata}
\end{center}
\end{figure*}

\begin{thebibliography}{99}
\bibitem{foot} Throughout this paper, a charge conjugate processes are assumed
implicitly.
\end{thebibliography}
\end{document}